\documentclass[12pt]{revtex4}    
\usepackage{amssymb,epsf}
\usepackage{latexsym}
\begin{document}
\title{Generalized second law of thermodynamics in Gauss-Bonnet braneworld}
\author{Ahmad Sheykhi $^{1,2}$\footnote{sheykhi@mail.uk.ac.ir} and Bin Wang $^{3}$\footnote{wangb@fudan.edu.cn}}
\address{$^1$Department of Physics, Shahid Bahonar University, P.O. Box 76175, kerman, Iran\\
         $^2$Research Institute for Astronomy and Astrophysics of Maragha (RIAAM), Maragha,
         Iran\\
$^3$  Department of Physics, Fudan University, Shanghai 200433,
China}
\begin{abstract}
We investigate the validity the generalized second law of
thermodynamics in a general braneworld model with curvature
correction terms on the brane and in the bulk, respectively.
Employing the derived entropy expression associated with the
apparent horizon, we examine the time evolution of the total
entropy, including the derived entropy of the apparent horizon and
the entropy of the matter fields inside the apparent horizon. We
show that the generalized second law of thermodynamics is fulfilled
on the $3$-brane embedded in the $5$D spacetime with curvature
corrections.

\end{abstract}
 \maketitle
It was first pointed out in \cite{Jac} that the hyperbolic second
order partial differential Einstein equation has a predisposition
to the first law of thermodynamics. This profound connection
between the first law of thermodynamics and the gravitational
field equations has been extensively observed in various gravity
theories \cite{Elin,Cai1,Pad}. Recently the study on the
connection between thermodynamics and gravity has been generalized
to the cosmological situations \cite{Cai2,Cai3,CaiKim,Fro,Wang},
where it was shown that the differential form of the Friedmann
equation on the apparent horizon in the FRW universe can be
rewritten in the form of the first law of thermodynamics. The
extension of this connection has also been carried out in the
braneworld cosmology \cite{Cai4,Shey1,Shey2}. The deep connection
between the gravitational equation describing the gravity in the
bulk and the first law of thermodynamics on the apparent horizon
reflects some deep ideas of holography.

Besides examining the validity of the thermodynamical
interpretation of gravity by expressing the gravitational field
equations into the first law of thermodynamics on the apparent
horizon in different spacetimes, it is of great interest to
examine other thermodynamical principles if the thermodynamical
interpretation of gravity really holds and is a generic feature.
This is especially interesting in the braneworld. In the
braneworld, the gravity is no longer Einstein gravity so that the
horizon entropy does not satisfy the usual area law. The entropy
on the apparent horizon in the braneworld was extracted through
connecting the gravitational equation to the first law of
thermodynamics on the apparent horizon \cite{Shey1,Shey2}. Whether
this obtained entropy satisfies general thermodynamical principles
is a question to be asked. In this work we are going to
investigate the generalized second law of thermodynamics by
examining the evolution of the apparent horizon entropy deduced
through the connection between gravity and the first law of
thermodynamics together with the matter fields' entropy inside the
apparent horizon. If the thermodynamical interpretation of gravity
is correct, the extracted apparent horizon entropy from the
profound connection between gravitational equations and the first
law of thermodynamics should satisfy the generalized second law of
thermodynamics.

The generalized second law of thermodynamics is an important
principle in governing the nature. Recently the generalized second
law in the accelerating universe enveloped by the apparent horizon
has been investigated in \cite{wang1,wang2}. Using the general
expression of temperature at apparent horizon of FRW universe, it
has been shown that the generalized second law holds in Einstein,
Gauss-Bonnet and more general lovelock gravity \cite{akbar}. Other
studies on the generalized second law of thermodynamics have been
done in \cite{Pavon2,Other, Setare}. Here we will consider a
general braneworld models with correction terms, such as a 4D
scalar curvature from induced gravity on the brane, and a 5D
Gauss-Bonnet curvature term in the bulk. With these correction
terms, especially including a Gauss-Bonnet correction to the 5D
action, we have the most general action with second-order field
equations in 5D \cite{lovelock}, which provides the most general
model for the braneworld scenarios \cite{RS,DGP}. In an effective
action approach to the string theory, the Gauss-Bonnet term
corresponds to the leading order quantum corrections to gravity,
its presence guarantees a ghost-free action\cite{zwiebach}. We
will first adopt the method developed in \cite{Shey2} to extract
the entropy expression associated with the apparent horizon from
the first law of thermodynamics in Gauss-Bonnet braneworld. We
will check the time evolution of the total entropy, including  the
entropy of the apparent horizon together with the entropy of the
matter fields inside the apparent horizon.

We start with the following action
\begin{eqnarray}\label{Act}
S  &=&  \frac{1}{2{\kappa_5}^2} \int{
d^5x\sqrt{-{g}}\left({R}-2\Lambda+\alpha
\mathcal{L}_{GB}\right)} \nonumber\\
 && + \frac{1}{2{\kappa_4}^2}\int
{d^4x\sqrt{-\widetilde{g}}\widetilde{R}}+\int
{d^{4}x\sqrt{-\widetilde{g}}( \mathcal {L}_{m} -2 \lambda)},
\nonumber
\end{eqnarray}
where $\Lambda<0$ is the bulk cosmological constant and
\begin{eqnarray}
{\mathcal L}_{GB}=R^2-4R^{AB}R_{AB}+R^{ABCD}R_{ABCD},
\end{eqnarray}
is the Gauss-Bonnet correction term. Quantities on the brane are
expressed with tilde to be distinguished from those in the bulk.
$\kappa_4$ and $\kappa_5 $ are the gravitational constants on the
brane and in the bulk and can be written respectively as
\begin{equation}\label{rela}
\kappa_{4}^2=8\pi G_{4}, \   \   \kappa_{5}^2=8\pi G_{5}.
\end{equation}
The last term in the action corresponds to the matter content.
$\mathcal{L}_m$ is the Lagrangian density of the brane matter
fields, and $\lambda$ is the brane tension. Hereafter we assume
that the brane cosmological constant is zero (if it does not
vanish, one can absorb it to the stress-energy tensor of matter
fields on the brane). We assume that there are no sources in the
bulk other than $\Lambda$.

The field equations can be obtained by varying the action with
respect to the bulk metric $g_{AB}$. The result is
\begin{equation}
  G_{AB}+\Lambda g_{AB}+2\alpha
H_{AB} = {\kappa_5}^2 T_{AB}\delta(y), \nonumber
 \end{equation}
where we have assumed that the brane is located at $y=0$ with $y$
being the coordinate in the extra dimension, and
\begin{equation}
H_{AB}=RR_{AB}-2R_A{}^CR_{BC}-2R^{CD}R_{ACBD}+R_A{}^{CDE}R_{BCDE}-\textstyle{1\over4}g_{AB}{\cal
L}_{GB},
 \end{equation}
is the second-order Lovelock tensor. The energy-momentum tensor
can be decomposed into two parts ${T}_{AB}=
\widetilde{T}_{AB}-\frac{1}{{\kappa_4}^2}
 \widetilde{G}_{AB}$,
where $\widetilde{T}_{AB}$ is the energy-momentum tensor
describing the matter confined to the 3-brane which we assume in
the form of a perfect fluid, $ \widetilde{T}_{AB} = (\rho+P)u_A
u_B + P\widetilde{g}_{AB} $, in the homogenous and isotropic
universe on the brane, where $u^A$, $\rho$, and $P$, are the fluid
velocity ($u^A u_A=-1$), energy density and pressure respectively.
The possible contribution of nonzero brane tension $\lambda$ will
be assumed to be included in $\rho$ and $P$. A homogeneous and
isotropic brane at fixed coordinate position $y=0$ in the bulk is
described by the line element
\begin{equation}
ds^2=-N^2(t,y) dt^2 + A^2(t,y)\gamma_{ij}dx^i dx^j
 + B^2(t,y)dy^2,  \nonumber
 \end{equation}
where $\gamma _{ij}$ is a maximally symmetric $3$-dimensional
metric for the surface ($t$=const., $y$=const.), whose spatial
curvature is parameterized by $k = -1, 0, 1$. On every
hypersurface ($y$=const), we have the metric of a FRW cosmological
model. The metric coefficient $N$ is chosen so that, $N(t,0)=1$
and $t$ is the cosmic time on the brane. The presence of the
four-dimensional curvature scalar in the gravitational action does
not affect the bulk equations. If we define
\begin{equation}
\Phi=\frac{1}{N^{2}}\frac{\dot{A}^{2}}{A^{2}}-\frac{1}{B^{2}}
\frac{A^{\prime \,2}}{A^{2}}+\frac{k}{A^{2}}
\end{equation}
 then the field equation reduces to \cite{germani}
\begin{eqnarray}\label{Field1}
&& \frac{\dot{A}}{A}\frac{N'}{N}+\frac{A'}{A}\frac{\dot{B}}{B}-
\frac{\dot{A}'}{A}=0,  \nonumber \\
&&\Phi+2\alpha\Phi^{2}+\frac{1}{\ell^2}-\frac{\mathcal{C}}{A^{4}}=0,\label{Field2}
\end{eqnarray}
where $\ell^2\equiv-6/\Lambda$ and $\mathcal{C}$ is an integration
constant which is related to the mass of the bulk black hole. In
the above equations dot and prime denote derivatives with respect
to $t$ and $y$, respectively. Integrating the $(00)$ component of
the field equation across the brane and imposing $\mathbb{Z}_2$
symmetry, we have the jump across the brane \cite{kofin}
\begin{eqnarray}\label{Jc1}
&&\frac{2{\kappa_4}^2}{{\kappa_5}^2}\left[
1+4\alpha\left(H^2+\frac{k}{a^{2} }- \frac{A^{\prime
\,2}_{+}}{3a^2b^2 } \right)\right]\frac{A'_{+}}{ ab }
 =-\frac{\kappa^{2}_{4}}{3}\rho+H^2 +\frac{k}{a^{2}},
\end{eqnarray}
where $a=A(t,0) $, $b=B(t,0) $ and $ 2A'_{+}=-2A'_{-}$ is the
discontinuity of the first derivative. $H=\dot a/a$ is the Hubble
parameter on the brane. Inserting $\Phi$ into Eq. (\ref{Jc1}) we
can obtain the generalized Friedmann equation on the
brane~\cite{kofin}
 \begin{eqnarray}\label{GFri}
&&\epsilon\frac{2{\kappa_4}^2}{{\kappa_5}^2}\left[1
+\frac{8}{3}\alpha\left(H^2 +{k \over a^2} + {\Phi_{0}\over 2}
\right) \right]\left(H^2 +{k \over a^2}-\Phi_{0}\right)^{1/2}
=-\frac{\kappa^{2}_{4}} {3}\rho+H^2 +{k \over a^2} ,
\label{3fried}
 \end{eqnarray}
where $\Phi_0=\Phi(t,0)$ and $\epsilon=\pm1$. For later
convenience we choose $\epsilon=-1$.  This Friedmann equation
contains various special cases discussed extensively in the
literature. The DGP braneworld is the limiting case when
$\alpha=0$, while the RS II braneworld can be reproduced in the
limit $\kappa_{4}\to \infty$ and $\alpha=0$. The pure Gauss-Bonnet
braneworld is the case with $\kappa_4 \to \infty$. To have further
understanding about the nature of the apparent horizon we rewrite
more explicitly, the metric of homogenous and isotropic FRW
universe on the brane in the form
\begin{equation}
ds^2={h}_{\mu \nu}dx^{\mu} dx^{\nu}+\tilde{r}^2d{\Omega_{2}}^2,
\end{equation}
where $\tilde{r}=a(t)r$, $x^0=t, x^1=r$, the two-dimensional
metric $h_{\mu \nu}$=diag $(-1, a^2/(1-kr^2))$, and $d\Omega_{2}$
is the metric of two-dimensional unit sphere. Then, the dynamical
apparent horizon, a marginally trapped surface with vanishing
expansion, is determined by the relation $h^{\mu
\nu}\partial_{\mu}\tilde {r}\partial_{\nu}\tilde {r}=0$, which
implies that the vector $\nabla \tilde {r}$ is null on the
apparent horizon surface. The apparent horizon has been argued to
be a causal horizon for a dynamical spacetime and is associated
with gravitational entropy and surface gravity \cite{Hay2,Bak}.
The explicit evaluation of the apparent horizon for the FRW
universe gives the apparent horizon radius
\begin{equation}
\label{radius}
 \tilde{r}_A=\frac{1}{\sqrt{H^2+k/a^2}}.
\end{equation}
The associated surface gravity on the apparent horizon can be
defined as
\begin{equation}
\label{surgra}\label{kappa}
 \kappa =\frac{1}{\sqrt{-h}}\partial_{a}\left(\sqrt{-h}h^{ab}\partial_{ab}\tilde
 {r}\right).
\end{equation}
Then one can easily show that the surface gravity at the apparent
horizon of FRW universe can be written as
\begin{equation}\label{surgrav}
\kappa=-\frac{1}{\tilde r_A}\left(1-\frac{\dot {\tilde
r}_A}{2H\tilde r_A}\right).
\end{equation}
The associated temperature on the apparent horizon can then be
defined as
\begin{equation}\label{Therm}
T_{h} =\frac{|\kappa|}{2\pi}=\frac{1}{2\pi \tilde
r_A}\left(1-\frac{\dot {\tilde r}_A}{2H\tilde r_A}\right).
\end{equation}
where $\frac{\dot{\tilde r}_A}{2H\tilde r_A}<1$ ensures that the
temperature is positive. Recently the connection between temperature
on the apparent horizon and the Hawking radiation has been observed
in \cite{cao}. Hawking radiation is an important quantum phenomenon
of black hole, which is closely related to the existence of event
horizon of black hole.  The cosmological event horizon of de Sitter
space has the Hawking radiation with thermal spectrum as well.
Using the tunneling approach proposed by Parikh and Wilczek, the
authors of \cite{cao} showed that there is indeed a Hawking
radiation with a finite temperature, for locally defined apparent
horizon of Friedmann-Robertson-Walker universe with any spatial
curvature. This gives more solid physical implication of the
temperature associated with the apparent horizon.

Now we turn to apply these results to investigate the validity of
the generalized second law of thermodynamics in the Gauss-Bonnet
braneworld scenario. We apply the approach we developed in
\cite{Shey1} to find the expression of entropy associated with the
horizon geometry in Gauss-Bonnet braneworld. First of all, we
assume that there is no black hole in the bulk and thus
$\mathcal{C}=0$. Inserting this condition into Eq. (\ref{Field2}),
we get $
 \Phi=\Phi_0=\frac{1}{4\alpha}\left(-1+\sqrt{1-\frac{8\alpha}{\ell^2}}\right)=\rm {const}.
$ This condition also implies $ \alpha<\ell^2/8$.  In terms of the
apparent horizon radius, we can rewrite the Friedmann equation
(\ref{GFri}) into
\begin{eqnarray}
\label{Fri1}
 \rho&=&\frac{3}{8\pi G_4}\frac{1}{{\tilde {r}_A}^2}+
\frac{3}{4\pi G_5} \left(\frac{1}{{\tilde {r}_{A}}^2}
-\Phi_{0}\right)^{1/2} \times\left[1
+\frac{8}{3}\alpha\left(\frac{1}{{\tilde {r}_{A}}^2} +
{\Phi_{0}\over 2} \right) \right].
 \end{eqnarray}
where we have used Eq. (\ref{rela}). Now, differentiating Eq.
(\ref{Fri1}) with respect to the cosmic time and using the
continuity equation
\begin{equation}
\label{Cont}
 \dot{\rho}+3H(\rho+P)=0,
\end{equation}
we get
\begin{equation} \label{dotr1}
\dot{\tilde{r}}_{A}=4\pi
H(\rho+P){\tilde{r}_{A}^2}\left[\frac{1}{G_{4}\tilde{r}_{A}}
+\frac{1}{G_{5}\sqrt{1-\Phi_0 \tilde {r}_{A}^2}}+\frac{4\alpha}{
G_{5}\tilde{r}_{A}^2}\left(\frac{2-\Phi_0 \tilde {r}_{A}^2
}{\sqrt{1-\Phi_0 \tilde {r}_{A}^2}}\right)\right]^{-1}.
\end{equation}
One can see from the above equation that $\dot{\tilde{r}}_{A}>0$
provided that the dominant energy condition, $\rho+P>0$, holds. In
our previous work \cite{Shey2}, we showed that the Friedmann
equation in a general braneworld model with curvature correction
terms can be written in the form of the first law of
thermodynamics on the apparent horizon of the brane,
\begin{equation}
dE=T_h dS_{h}+WdV,
\end{equation}
where $W=(\rho-P)/2$ is the matter work density \cite{Hay2} which
is regarded as the work done by the change of the apparent
horizon, $E=\rho V$ is the total energy of the matter fields on
the brane inside a 3-ball with the volume
$V=\Omega_3\tilde{r}_{A}^{3}$. The area of the apparent horizon is
the area of the ball on a 2-sphere of radius $\tilde{r}_{A}$,
which is expressed as $A=3\Omega_3\tilde{r}_{A}^2$ \cite{TT}.
Using the first law we can extract the expression for the entropy
on the apparent horizon in the general Gauss-Bonnet braneworld
\cite{Shey2}
\begin{eqnarray}\label{ent1}
 S_{h} &=& \frac{3\Omega_{3}}{2G_{4}}{\displaystyle\int^{\tilde
r_A}_0\tilde{r}_{A}d\tilde {r}_{A}}+
\frac{3\Omega_{3}}{2G_{5}}{\displaystyle\int^{\tilde r_A}_0
\frac{\tilde{r}_{A}^{2}d\tilde {r}_{A}}{\sqrt{1-\Phi_0 \tilde
{r}_{A}^2}}}\nonumber \\
&+& \frac{6\alpha\Omega_{3}}{G_{5}}{\displaystyle\int^{\tilde
r_A}_0 \frac{2-\Phi_0 \tilde {r}_{A}^2 }{\sqrt{1-\Phi_0 \tilde
{r}_{A}^2}}d\tilde {r}_{A}}.
\end{eqnarray}
The explicit form of the entropy can be obtained by integrating
(\ref{ent1}), which reads
\begin{eqnarray} \label{ent2}
S_{h}
&=&\frac{3\Omega_{3}{\tilde{r}_A}^{2}}{4G_{4}}+\frac{2\Omega_{3}{\tilde{r}_A}^{3}}{4G_{5}}
\times {}_2F_1\left(\frac{3}{2},\frac{1}{2},\frac{5}{2},
\Phi_0{\tilde{r}_A}^2\right) \nonumber \\
 && +\frac{6\alpha\Omega_{3}{\tilde{r}_A}^3}{G_{5}} \left[
\Phi_0 \times {}_2F_1\left(\frac{3}{2},\frac{1}{2},\frac{5}{2},
\Phi_0{\tilde{r}_A}^2\right) +\frac{\sqrt{1-\Phi_0 \tilde
{r}_{A}^2}}{{\tilde{r}_A}^2}\right],
\end{eqnarray}
where ${}_2F_1(a,b,c,z)$ is a hypergeometric function. The
expression looks complicated. But its physical meaning is clear
and includes various special cases \cite{Shey2}. Some remarks are
as follows. The first term in (\ref{ent2}) has the form of
Bekenstein-Hawking entropy of the 4D Einstein gravity. In fact,
the Einstein-Hilbert term ($\tilde R$) on the brane contributes to
this area term. The second term is due to the Einstein-Hilbert
term ($R$) in the bulk. The Gauss-Bonnet correction term in the
bulk contributes to the last term. In the limit $\alpha\rightarrow
0$, this expression reduces to the entropy of apparent horizon in
warped DGP braneworld embedded in a $AdS_5$ bulk, while in the
limit $\alpha \rightarrow 0$ and $\Phi_0 \rightarrow 0$ (or
equality $\ell \rightarrow \infty$), it reduces to the entropy of
apparent horizon in pure DGP braneworld with a Minkowskian bulk
\cite{Shey1}. Furthermore, taking the limit $\alpha \rightarrow 0$
and  $G_4 \rightarrow \infty $, while keeping $G_5$ finite, the
first and the last terms in (\ref{ent2}) vanish and we obtain the
entropy associated with the apparent horizon in RSII braneworld
\cite{Shey1}. Finally, keeping $\alpha$ finite, and taking the
limit $G_4 \rightarrow \infty $ and  $\Phi_0 \rightarrow 0$, one
can extract from Eq. (\ref{ent2}) the entropy associated with the
apparent horizon on the brane in the Gauss-Bonnet braneworld with
a Minkowskian bulk
\begin{eqnarray}
 S_{h}=\frac{2\Omega_{3}{\tilde{r}_A}^{3}}{4G_{5}}
 \left(1+\frac{12 \alpha}{{\tilde{r}_A}^{2}}\right).
\end{eqnarray}
This gives an expression of horizon entropy in the Gauss-Bonnet
gravity~\cite{GB,Cai2,Cai3,CaiKim,Cvetic}. This is an expected
result. Indeed, because of the absence of scalar curvature term on
the brane and the negative cosmological constant in the bulk, no
localization of gravity happens on the brane. As a result, the
gravity on the brane is still $5$D Gauss-Bonnet gravity and the
brane looks like a domain wall moving in a Minkowski spacetime.
Therefore, the entropy of apparent horizon on the brane still
obeys the entropy formula in the bulk. Note that the factor $2$ in
the entropy comes from the $\mathbb{Z}_2$ symmetry of the bulk.
These results give a self-consistency check for the entropy
expression (\ref{ent2}).

Next we turn to find out $T_{h} \dot{S_{h}}$. It is a matter of
calculation to show that
\begin{equation}\label{TSh}
T_{h} \dot{S_{h}} =\frac{3\Omega_{3}}{4\pi}\left(1-\frac{\dot
{\tilde r}_A}{2H\tilde r_A}\right)\left[\frac{1}{G_{4}}+
\frac{\tilde{r}_{A}}{G_{5}\sqrt{1-\Phi_0 \tilde {r}_{A}^2}}+
\frac{4\alpha}{G_{5}\tilde{r}_{A}}\left(\frac{2-\Phi_0 \tilde
{r}_{A}^2 }{\sqrt{1-\Phi_0 \tilde
{r}_{A}^2}}\right)\right]\dot{\tilde{r}_{A}}.
\end{equation}
Substituting Eq. (\ref{dotr1}) and doing some simplifications, we
obtain
\begin{equation}\label{TSh2}
T_{h} \dot{S_{h}} =3\Omega_{3} H(\rho+P){\tilde
r_A}^{3}\left(1-\frac{\dot {\tilde r}_A}{2H\tilde r_A}\right).
\end{equation}
As we argued above that the physical positive temperature $T_h$
requires the term $\left(1-\frac{\dot {\tilde r}_A}{2H\tilde
r_A}\right)$ to be positive.  However, in the accelerating
universe the dominant energy condition may be violated,
$\rho+P<0$, which can make the second law of thermodynamics
,$\dot{S_{h}}\geq0$, do not hold in general. Then the question
arises, ``will the generalized second law of thermodynamics,
$\dot{S_{h}}+\dot{S_{m}}\geq0$, still holds on the brane?''
According to the generalized second law the entropy of matter
fields inside the horizon together with the entropy on the
apparent horizon cannot decrease with time. The entropy of the
matter fields inside the apparent horizon, $S_{m}$, can be related
to its energy $E=\rho V$ and pressure $P$ in the horizon by the
Gibbs equation \cite{Pavon2}
\begin{equation}\label{Gib}
T_m dS_{m}=d(\rho V)+PdV=V d\rho+(\rho+P)dV,
\end{equation}
where $T_{m}$ is the temperature of the energy inside the horizon.
We limit ourselves to the assumption that the thermal system
enveloped by the apparent horizon remains in equilibrium so that
the temperature of the system must be uniform and the same as the
temperature of its boundary. This requires that the temperature
$T_m$ of the energy inside the apparent horizon should be in
equilibrium with the temperature $T_h$ associated with the
apparent horizon, so we have $T_m = T_h$. This expression holds in
the local equilibrium hypothesis. If the temperature of the fluid
differs much from that of the horizon, there will be spontaneous
heat flow between the horizon and the fluid inside and the local
equilibrium hypothesis will no longer hold. Therefore from the
Gibbs equation (\ref{Gib}) we can obtain
\begin{equation}\label{TSm2}
T_{h} \dot{S_{m}} =(n-1)\Omega_{n-1} {\tilde r_A}^{n-2}\dot
{\tilde r}_A(\rho+P)-(n-1)\Omega_{n-1}{\tilde r_A}^{n-1}H(\rho+P).
\end{equation}
To check the generalized second law of thermodynamics, we have to
examine the evolution of the total entropy $S_h + S_m$. Adding
equations (\ref{TSh2}) and (\ref{TSm2}),  we get
\begin{equation}\label{GSL1}
T_{h}( \dot{S_{h}}+\dot{S_{m}})=\frac{3\Omega_{3}}{2}{\tilde
r_A}^{2}\dot {\tilde r}_A(\rho+P)=\frac{A}{2}(\rho+P) \dot {\tilde
r}_A.
\end{equation}
Substituting $\dot {\tilde r}_A$ from Eq. (\ref{dotr1}) into
(\ref{GSL1}) we get
\begin{equation}\label{GSL11}
T_{h}( \dot{S_{h}}+\dot{S_{m}})=2\pi A {\tilde r_A}^{2}
H(\rho+P)^2\left[\frac{1}{G_{4}\tilde{r}_{A}}
+\frac{1}{G_{5}\sqrt{1-\Phi_0 \tilde {r}_{A}^2}}+\frac{4\alpha}{
G_{5}\tilde{r}_{A}^2}\left(\frac{2-\Phi_0 \tilde {r}_{A}^2
}{\sqrt{1-\Phi_0 \tilde {r}_{A}^2}}\right)\right]^{-1}.
\end{equation}
It is worth noticing that $1-\Phi_0 \tilde {r}_{A}^2>0$, therefore
the right hand side of the above equation cannot be negative
throughout the history of the universe. Hence we have $
\dot{S_{h}}+\dot{S_{m}}\geq0$ indicating that the generalized
second law of thermodynamics is fulfilled in a general braneworld
model with curvature correction terms in a region enclosed by the
apparent horizon.

In summary, we have investigated the validity of the generalized
second law of thermodynamics in a general braneworld model with
curvature correction terms, such as a 4D scalar curvature from
induced gravity on the brane, and a 5D Gauss-Bonnet curvature term
in the bulk. Following the method developed in \cite{Shey2} we
have extracted the entropy associated with the apparent horizon in
a general Gauss-Bonnet braneworld by relating the gravitational
equation to the first law of thermodynamics on the apparent
horizon. We have examined the total entropy evolution with time,
including the derived apparent horizon entropy and the entropy of
matter fields inside the apparent horizon on the brane. We have
shown that the generalized second law of thermodynamics on the
$3$-brane embedded in the $5$D spacetime with curvature correction
terms is fulfilled throughout the history of the universe. The
satisfaction of the generalized second law of thermodynamics
provides further confidence on the thermodynamical interpretation
of gravity.

\acknowledgments{This work has been supported financially by
Research Institute for Astronomy and Astrophysics of Maragha,
Iran. The work of B. W. was partially supported by NNSF of China
and Shanghai Science and Technology Commission and Shanghai
Education Commission.}


\begin{thebibliography}{99}
\bibitem{Jac} T. Jacobson, Phys. Rev. Lett. {\bf75}, 1260 (1995).

\bibitem{Elin} C. Eling, R. Guedens, and T. Jacobson,
Phys. Rev. Lett. {\bf96}, 121301 (2006).
\bibitem{Cai1} M. Akbar and R. G. Cai, Phys. Lett. B {\bf635}, 7 (2006)
;\\ M.~Akbar and R.~G.~Cai,
   Phys. Lett. B {\bf648}, 243 (2007).

\bibitem{Pad}T. Padmanabhan, Class. Quant. Grav. {\bf 19}, 5387
(2002); \\ T. Padmanabhan, Phys. Rept. {\bf 406}, 49 (2005);\\
T.~Padmanabhan,
  gr-qc/0606061;\\
 A.~Paranjape, S.~Sarkar and T.~Padmanabhan,
    Phys.\ Rev.\ D {\bf 74}, 104015 (2006);\\
  D.~Kothawala, S.~Sarkar and T.~Padmanabhan,
   gr-qc/0701002;\\ T.~Padmanabhan and A.~Paranjape,
  Phys. Rev. D {\bf75} (2007) 064004;\\ S. F. Wu, B. Wang, and G. H Yang, Nucl. Phys. B {\bf799} (2008) 330.

  \bibitem{Cai2} M.~Akbar and R.~G.~Cai, Phys. Rev. D {\bf 75}, 084003 (2007).
  \bibitem{Cai3} R.~G.~Cai and L.~M.~Cao, Phys.Rev. D {\bf 75}, 064008
  (2007).

\bibitem{CaiKim} R. G. Cai and S. P. Kim, JHEP {\bf0502}, 050
(2005).

 \bibitem{Fro} A. V. Frolov and L. Kofman, JCAP {\bf 0305},
009 (2003);\\ U. K. Danielsson, Phys. Rev. D {\bf71}, 023516(2005)
;\\ R. Bousso, Phys. Rev. D {\bf71}, 064024 (2005);\\ G. Calcagni,
JHEP {\bf0509}, 060 (2005).
\bibitem{Wang} B. Wang, E.
Abdalla and R. K. Su, Phys.Lett. B {\bf503},  394 (2001);\\ B.
Wang, E. Abdalla and R. K. Su, Mod. Phys. Lett. A {\bf17},  23
(2002);\\ R.~G.~Cai and Y.~S.~Myung, Phys.\ Rev.\ D {\bf 67},
124021 (2003).
\bibitem{Cai4} R.~G.~Cai and L.~M.~Cao,
  Nucl. Phys. B {\bf785} (2007) 135.

  \bibitem{Shey1} A. Sheykhi, B. Wang and R. G. Cai, Nucl. Phys. B {\bf
779} (2007)1.
  \bibitem{Shey2} A. Sheykhi, B. Wang and R. G. Cai, Phys. Rev. D {\bf
76} (2007) 023515.

\bibitem{wang1} Bin Wang, Yungui Gong, Elcio Abdalla, Phys.Rev. D {\bf74} (2006) 083520.

\bibitem{wang2} Jia Zhou, Bin Wang, Yungui Gong, Elcio Abdalla,  Phys. Lett. B {\bf652} (2007) 86.

\bibitem{akbar} M. Akbar, gr-qc/0808.0169; \\ M. Akbar,
gr-qc/0808.3308.
\bibitem{Pavon2} G. Izquierdo and D. Pavon, Phys. Lett. B {\bf633} (2006)
420.
\bibitem{Other} E. Babichev, V. Dokuchaev, Yu. Eroshenko, Phys. Rev. Lett.
{\bf93} (2004) 021102;
\\ M. D. Pollock, T. P. Singh, Class.
Quantum Grav. {\bf6} (1989) 901;\\ P. C. W. Davies, Class. Quantum
Grav. {\bf4} (1987) L225;\\ Izquierdo, D. Pavon, Phys. Lett. B
{\bf639} (2006) 1.

\bibitem{Setare} M. R. Setare and S. Shafei, JCAP {\bf09} (2006) 011;\\ M. R.
Setare, Phys. Lett. B {\bf641} (2006) 130;\\ H. Mohseni Sadjadi,
Phys. Rev. D {\bf73} (2006) 063525; \\ H. Mohseni Sadjadi, Phys. Rev. D {\bf76} (2007) 104024;\\
 H. Mohseni
Sadjadi, Phys. Lett. B {\bf645} (2007) 108; \\ S. F. Wu, B. Wang,
G. H. Yang and P. M. Zhang, arXive: 0801.2688.


\bibitem{lovelock}
D. Lovelock, J. Math. Phys. {\bf 12}, 498 (1971).
\bibitem{RS} L. Randall, R. Sundrum, Phys. Rev. Lett. {\bf83} (1999)
3370;\\ L. Randall, R. Sundrum, Phys. Rev. Lett. {\bf 83} (1999)
4690.

\bibitem{DGP} G. Dvali, G. Gabadadze, M. Porrati, Phys. Lett. B {\bf 485}, 208 (2000)
;\\ G. Dvali, G. Gabadadze, Phys. Rev. D {\bf 63} 065007 (2001).

\bibitem{zwiebach}
B. Zwiebach, Phys. Lett. B {\bf 156}, 315 (1985); \\ B. Zumino,
Phys. Rept. {\bf 137}, 109 (1986);\\ D. J. Gross and J.H. Sloan,
Nucl. Phys. {\bf B291}, 41 (1987).

\bibitem{cao}  R.G. Cai, L.M. Cao, Y.P. Hu, arXiv:0809.1554; \\
 R. Li, J. R. Ren, D. F. Shi, Phys. Lett. B {\bf670} (2009) 446.

\bibitem{germani}
C. Germani and C. Sopuerta, Phys. Rev. Lett. {\bf 88}, 231101
(2002).
\bibitem{kofin}   G. Kofinas, R. Maartens,  E. Papantonopoulos, JHEP {\bf0310},  066 (2003)
.
\bibitem{Hay2} S. A. Hayward, S.Mukohyana, andM. C. Ashworth, Phys.
Lett.  A {\bf 256}, 347 (1999);\\ S. A. Hayward, Class. Quantum
Grav. {\bf 15}, 3147 (1998).
\bibitem{Bak} D. Bak and S. J. Rey, Class. Quantum Grav. {\bf17}, L83 (2000).
\bibitem{TT} $\Omega_{3}=\frac{\pi^{3/2}}{\Gamma(5/2)}=4\pi/3$.
For the benefit to generalize our results to the $(n-1)$-brane
embedded in the $(n+1)$-dimensional spacetime, we use $\Omega_3$
here.
\bibitem{GB}R.~C.~Myers and J.~Z.~Simon,
  Phys.\ Rev.\ D {\bf 38}, 2434 (1988);\\
R.~G.~Cai,
  Phys.\ Rev.\ D {\bf 65}, 084014 (2002)
  ;\\
R.~G.~Cai and Q.~Guo,
  Phys.\ Rev.\ D {\bf 69}, 104025 (2004).
 \bibitem{Cvetic} M. Cvetic, S. Nojiri, S. D. Odintsov, Nucl. Phys. B {\bf 628}, 295,
(2002). 
\end{thebibliography}
\end{document}